\documentclass{article}
\usepackage{spconf}
\usepackage{IEEEtrantools}
\usepackage{cite}
\usepackage{hyperref}
\usepackage{graphicx}
\usepackage{tikz}
	\usetikzlibrary{arrows.meta}
	\usetikzlibrary{shapes}
\usepackage{pgfplots}
	\usepgfplotslibrary{groupplots}
	\pgfplotsset{compat=newest,every axis/.append style={font=\footnotesize}}
\usepackage{amsmath,amssymb}
\usepackage{url}
\usepackage{algorithmic}
\usepackage{array}

\title{CLOCK SYNCHRONIZATION OVER NETWORKS USING SAWTOOTH MODELS}

\name{Pol del Aguila Pla$\dagger$, Lissy Pellaco$\ddagger$, Satyam Dwivedi$\dagger\dagger$, Peter H\"{a}ndel$\ddagger$, and Joakim Jald\'{e}n$\ddagger$\thanks{The first author performed the work while at the KTH Royal Institute of Technology. Contact authors, \texttt{pol.delaguilapla@epfl.ch} and \texttt{pellaco@kth.se}.} \thanks{This work was supported by the SRA ICT TNG project ``Privacy-preserved Internet Traffic Analytics'' (PITA).} }
\address{$\dagger$ Biomedical Imaging Group, EPFL, Lausanne, Switzerland \\ $\ddagger$ Division of Information Science and Engineering, School of EECS \\KTH Royal Institute of Technology, Stockholm, Sweden \\$\dagger\dagger$Ericsson Research, Stockholm, Sweden}

    \newcommand{\listintzero}[1]{\lbrace 0, 1, \dots, #1\rbrace}
	
    \newcommand{\mse}[1]{\operatorname{\mathrm{MSE}}( #1 )}
    \newcommand{\normal}[2]{\operatorname{\mathcal{N}}\!\left(#1,#2\right)}
    \newcommand{\fish}[2]{\operatorname{\mathrm{I}}_{#2}^{#1}}
    \newcommand{\crlbu}[1]{\operatorname{\mathrm{CRLB}}_\mathrm{u}\!\left(#1\right)}
    \def\parsu{\boldsymbol{\omega}}
    
    \def\tr{\operatorname{\mathrm{Tr}}}
    \newcommand{\trans}[1]{{#1}^\mathrm{T}}
    \newcommand{\Id}[1]{\mathbf{\mathrm{I}}_{#1}}
    \newcommand{\ones}[1]{\mathbf{1}_{#1}}
    
    \renewcommand{\mod}[1]{\operatorname{mod}_{#1}\!}
    \def\m2pi{\mod{2\pi}}
    \def\modu{\mod{1}}
    \newcommand{\extrema}[3]{\operatorname{#1}_{#2}\left\lbrace #3 \right\rbrace}
    \renewcommand{\max}[2]{\extrema{max}{#1}{#2}}
    \def\func{\operatorname{g}}
    \def\sign{\operatorname{sign}}
    
    \def\freqd{f_{\mathrm{d}}} 
    
    \def\master{\mathcal{M}} 
    \def\slave{\mathcal{S}}
    \def\samplingperiod{T_{\mathrm{s}}} 
    \def\slaveperiod{T_{\slave}} 
    \def\masterperiod{T_{\master}}
    \def\delay{\delta_0} 
    \def\delayping{\delta_{\rightarrow}}
    \def\delaypong{\delta_{\leftarrow}}
    \def\delaytrans{\delta_{\leftrightarrow}}
    \def\slavephase{\phi_{\slave}} 
    \def\masterphase{\phi_{\master}}
    \def\SNRin{\mathrm{SNR}_\mathrm{in}}
    \def\SNRout{\mathrm{SNR}_\mathrm{out}}

    \def\talpha{\tilde{\alpha}} \def\tbeta{\tilde{\beta}} \def\n{\mathbf{n}}

\begin{document}
    % Everything in nine points
    \ninept
    
    % Make title
    \maketitle
    
    % Abstract
    \begin{abstract}
        Clock synchronization and ranging over a wireless network with 
        low communication overhead is a challenging goal with tremendous 
        impact. In this paper, we study the use of time-to-digital converters
        in wireless sensors, which provides clock synchronization and ranging at 
        negligible communication overhead through a sawtooth signal model for round 
        trip times between two nodes. In particular, we derive Cram\'{e}r-Rao lower bounds for a 
        linearitzation of the sawtooth signal model, and we thoroughly
        evaluate simple estimation techniques by simulation, giving clear and concise 
        performance references for this technology.
    \end{abstract}
    
    % Keywords

    \section{Introduction}
    \label{sec:intro}
        Time-to-digital converters (TDC) are
        independently clocked, low-power, highly 
        accurate time measurement devices. 
        Incorporating TDCs in the design of wireless
        sensors provides very accurate ranging 
        information from basic round trip time (RTT) 
        measurement protocols~\cite{Angelis2013}.
        Such a scheme has been used to devise 
        reliable and cost-efficient systems for indoor 
        localization~\cite{Nilsson2014}. 
        A similar scheme, introduced in \cite{Dwivedi2015},
        uses an improved RTT protocol to address 
        clock synchronization across a deployed 
        network. This approach is extensively analyzed 
        in~\cite{AguilaPla2020a},
        both practically and theoretically. Clock
        synchronization becomes possible due to the presence 
        of two different clock speeds within each wireless sensor,
        i.e., that of the sensor and that of its TDC. 
        The resulting RTT measurements follow a 
        sawtooth signal model~\cite{Dwivedi2015}, which, under
        realistic assumptions, leads to the identifiability of
        the clock synchronization and ranging 
        parameters~\cite{AguilaPla2020a}.
        In this paper, we provide performance references for the
        use of this technology to synchronize two nodes in a wireless network, which will benefit both engineers 
        that use it and researchers studying the 
        estimation of sawtooth signal models.
        
        Clock synchronization in wireless sensor networks has been
        studied extensively from a variety of perspectives
        \cite{Freris2010,He2014,Etzlinger2014,Gholami2015,Kim2017,Zachariah2017,He2017,Etzlinger2018,Xia2018}. 
        Most studies focus on global synchronization
        performance through a network based on some form of time-stamped 
        message exchange. Some of these target specific objectives, e.g.,
        a fast consensus across the network \cite{He2014,He2017}
        or energy efficiency \cite{Kim2017,Zachariah2017}, but communication
        overhead due to the arguably unnecessary exchange of time stamps 
        is usually disregarded. However, several works
        \cite{Zachariah2014,Gholami2015,Dwivedi2015,Etzlinger2018,AguilaPla2020a}
        have reported that two-way message exchanges without time stamps
        have the potential to substantially lower communication overhead
        while still providing accurate synchronization. Our study provides
        performance references on the synchronization accuracy of two TDC-equipped sensors in 
        a WSN that exchange messages without time stamps, reducing 
        communication overhead and obtaining remarkable performance
        in ranging and frequency synchronization (errors under $0.1~\mathrm{cm}$
        and $1~\mathrm{ppb}$ of the clock frequency).

    \section{Sawtooth model and Cram\'er-Rao lower bounds}
    \label{sec:model} \label{sec:CRLBs}
        An empirical study run by our group~\cite{Dwivedi2015} revealed that
with a specific measurement protocol 
(see~\cite{Dwivedi2015}~and~\cite{AguilaPla2020a}), the RTTs $Y[n]$ 
measured between two sensors with TDCs, which we name $\master$ and $\slave$, follow the 
sawtooth signal model, i.e., 
\begin{IEEEeqnarray}{rl}\label{eq:model}
    Y[n] \,& = \alpha + W[n] + \psi \modu\left(\beta n + \gamma + V[n] \right)\,,
\end{IEEEeqnarray}
where $W[n]$ and $V[n]$ are noise processes, which are assumed to be white,
independent, zero-mean Gaussian processes with standard deviations $\sigma_v$
and $\sigma_w$, respectively. Here, $\alpha$, $\psi$, $\beta$ and $\gamma$ are the 
generic sawtooth model parameters, for offset, amplitude, normalized frequency,
and phase.
In~\cite{AguilaPla2020a}, we show that under simple modeling
assumptions, when $\master$ measures RTTs to and from $\slave$, it obtains
\begin{IEEEeqnarray}{rl} \label{eq:stoch_model}
    Y[n] \,& = \delaytrans + \delay + W[n] + \slaveperiod H[n] \mbox{, where } \\
    H[n] \,& = 1 - \mod{1}\left[ 
                        \samplingperiod \freqd n + \frac{\delayping}{\slaveperiod} 
                        + \frac{\slavephase}{2\pi} + V[n] 
                    \right]\,. \nonumber
\end{IEEEeqnarray}
Here, $\delay~[\mathrm{s}]$ is a known delay introduced by $\slave$, 
$\delaytrans \approx 2\rho/c~[\mathrm{s}]$ is the transmission time of each 
message back and forth, which we assume to be the result of two identic delays, $\delayping = \delaypong$, and where 
$\rho~[\mathrm{m}]$ is the range between $\master$ and $\slave$ and $c~[\mathrm{m/s}]$ is the speed of light in the communication medium. Further, 
$\slaveperiod~[\mathrm{s}]$ (unknown by $\master$) and $\masterperiod~[\mathrm{s}]$ (known by $\master$, measured through its TDC) are, respectively, the clock periods of $\slave$ and $\master$, while $\freqd = 1/\slaveperiod - 1/\masterperiod~[\mathrm{Hz}]$ is the difference between their frequencies,
and $\samplingperiod = K \masterperiod$ is the known time between two consecutive
measurements. Finally, $\slavephase~[\mathrm{rad}]$ is the unknown phase of $\slave$'s 
clock when $\masterphase=0~\mathrm{rad}$ is assumed.

            In~\cite{AguilaPla2020a}, we show that 
    \eqref{eq:stoch_model} is an identifiable model, i.e.,
    that the distribution of the data contains enough information
    to singularly identify these parameters.
    Nonetheless, the likelihood function is not 
    differentiable everywhere. This violates the 
    assumptions of the Cram\'er-Rao lower 
    bound (CRLB) for the mean square error (MSE) of
    unbiased estimators, hindering our objective 
    of providing performance references for the estimation
    of the model's parameters.
    Instead, we analyze a linear model that results from 
    assuming that an oracle has removed the effect of the
    nonlinearity (phase unwrapping). The model then becomes
    \begin{IEEEeqnarray}{c}
      \label{eq:unwrappedmodel}
      \!\!\!\!\!\!\!\!\!\!Z[n] = \delay + \frac{\delaytrans}{2} + \slaveperiod \left( 1- \frac{\slavephase}{2\pi}\right) - \slaveperiod \samplingperiod \freqd n  + U[n]\,,
    \end{IEEEeqnarray}
    with $U[n]$ a white Gaussian process such that $U[n]\sim\normal{0}{\sigma^2}$ with
    $\sigma^2 = \sigma_w^2+ \slaveperiod^2 \sigma_v^2$. The resulting model \eqref{eq:unwrappedmodel}
    is not without complications. First, $\slavephase$ and $\delaytrans$ are not jointly identifiable,
    because only their weighted sum affects the distribution
    of $Z[n]$. 
    Second, the variance of the noise now depends on
    $\slaveperiod$, i.e., on $\freqd$, one of the parameters to estimate. Therefore, we analyze first a general
    linear model with slope-dependent noise power, i.e., the model
    \begin{equation} \label{eq:vectorunwrappedmodel}
      \mathbf{Z} = \left[ \ones{}, \n  \right] \parsu %\left[ \talpha, \tbeta \right]^{\mathrm{T}}
                  + \mathbf{U} \mbox{, with }\mathbf{U}\sim\normal{0}{\sigma^2\Id{N}}\mbox{, with}
    \end{equation}  
    $\sigma^2 = \sigma_0^2 + ( \sigma_1 + \tbeta \sigma_2 )^2$, $\mathbf{Z}=\trans{\left[Z[0],Z[1],\dots,Z[N-1]\right]}$, and $\parsu = \trans{[\talpha,\tbeta]}$, and where $\ones{}$ and
    $\n$ are $N$-dimensional vectors with ones and the sorted indices between $0$ and $N-1$, respectively, and $\sigma_0\geq0$, $\sigma_1\geq0$ and $\sigma_2\geq0$ are known.
    This model is equivalent to \eqref{eq:unwrappedmodel} when $\talpha =\delay +\delaytrans/2+\slaveperiod(1-\slavephase/2\pi)$,
    $\tbeta =  - \slaveperiod \samplingperiod \freqd$, 
    $\sigma_0 = \sigma_w$, $\sigma_1 = \masterperiod \sigma_v$, and 
    $\sigma_2 = \sigma_v /K$. Here, recall that $K=\samplingperiod/\masterperiod$. The advantages of \eqref{eq:vectorunwrappedmodel} with respect to \eqref{eq:unwrappedmodel} are that i) it is an 
    identifiable model, and ii) it can be analyzed using standard results for the Fisher information matrix of Gaussian models~\cite[ch.~3.9, p.~47]{Kay1993}. Furthermore, 
    given the Fisher information matrix $\fish{}{\parsu}$ for \eqref{eq:vectorunwrappedmodel}, one can obtain CRLBs for $\freqd$, for $\slavephase$ when $\delaytrans$ is known,
    and for $\delaytrans$ when $\slavephase$ is known, by using the CRLB on functions of vector parameters~\cite[corollary~5.23, p.~306]{Schervish1995}, i.e., 
    \begin{IEEEeqnarray}{c}\label{eq:CRLB}
      \!\!\!\!\mse{\hat{\func}(\mathbf{Z})} \geq \crlbu{\func(\parsu)} = \trans{\left(\nabla_{\parsu}\func\right)} \fish{-1}{\parsu} \left(\nabla_{\parsu}\func\right),
    \end{IEEEeqnarray}
    where $\hat{\func}(\mathbf{Z})$ is an unbiased estimator of $\func(\parsu)$, a bounded function,
    and $\nabla_{\parsu}\func$ is its gradient.
    The derivation and statement of the inverse Fisher information matrix for \eqref{eq:CRLB} can be found in Section~\ref{sec:appFish}. 
    Then, from the relation between \eqref{eq:unwrappedmodel}~and~\eqref{eq:vectorunwrappedmodel}, one obtains that
    \begin{IEEEeqnarray}{lll} \label{eq:relation_parameters_sync}
      \freqd &= \func_{\freqd}(\parsu) &= -\frac{\tbeta}{\masterperiod\left(K \masterperiod + \tbeta\right)}\,, \\
      \slavephase & = \func_{\slavephase}(\parsu)  &= 2\pi + \frac{2\pi}{\masterperiod + \frac{\tbeta}{K}}\left( \frac{\delaytrans}{2}+\delay - \talpha\right)\,,\mbox{ and} \nonumber \\
      \delaytrans &=  \func_{\delaytrans}(\parsu) &= 2\left( \talpha - \delay - \left( \masterperiod + \frac{\tbeta}{K}\right)\left(1 -\frac{\slavephase}{2\pi}\right) \right)\,. \nonumber 
    \end{IEEEeqnarray}
    The expressions for $\slavephase$ or $\delaytrans$ assume that the respective other is known. This circumvents the joint identifiability problem stated above, but the resulting CRLBs will disregard that both parameters need to be estimated simultaneously.
    Nonetheless, our purpose in deriving these bounds is to use them as a plausible reference for the performance one can obtain
    using \eqref{eq:stoch_model}, for which we proved identifiability in~\cite{AguilaPla2020a}.
    In order to establish the CRLBs using \eqref{eq:CRLB} we obtain
    \begin{IEEEeqnarray}{ll}\label{eq:gradients}
      \nabla_{\parsu}\func_{\freqd}(\parsu) & = -\frac{1}{\slaveperiod^2K}\trans{[0,1]}\mbox{, } \\
      \nabla_{\parsu}\func_{\delaytrans}(\parsu) & = 2\trans{\left[1,\frac{1}{K}\left(\frac{\slavephase}{2\pi}-1\right)\right]}\mbox{, and } \nonumber \\
      \nabla_{\parsu}\func_{\slavephase}(\parsu) &= \frac{-2\pi}{\slaveperiod}\trans{\left[1,\frac{1}{K}\left(\frac{\slavephase}{2\pi}-1\right)\right]}. \nonumber 
    \end{IEEEeqnarray}
       
    The obtained CRLBs are valid for unbiased estimators from data $\mathbf{Z}$ generated according 
    to \eqref{eq:unwrappedmodel}, but they are not guaranteed to hold for unbiased estimators from
    data $\mathbf{Y}$ generated from \eqref{eq:stoch_model}. Furthermore, they are not valid bounds
    on the MSE of biased estimators from either model. Nonetheless, we believe they provide a
    linear intuition that, as our experimental results confirm, is practically relevant.

    \section{Basic estimation strategies} \label{sec:estimation}
            
    We present simple estimators for the parameters of a sawtooth signal model~\eqref{eq:model}
    based on the techniques proposed in~\cite{Dwivedi2015}. In their simplicity, they show remarkable robustness for the 
    ranges of parameters $\alpha$, $\beta$, $\gamma$ and $\psi$ that arise in practical clock synchronization and 
    ranging scenarios. Consequently, we consider them to be a good reference on the minimum expected performance that can
    be obtained from systems that use the proposed technology. For the sake of reproducibility and direct impact, we provide thoroughly documented Jupyter notebooks that contain the implementation of all the presented techniques in this project's repository~\cite{GITHUB}.
    
    We expose our estimation methods in the more general notation of~\eqref{eq:model}. However, we will consider that given 
    $\beta$ or $\psi$, the other is fully determined. This parallels clock synchronization, in which 
    $\beta = \samplingperiod \freqd$ and $\psi = -\slaveperiod = -\masterperiod/(\masterperiod \freqd + 1)$. 
    For practical application of these techniques to clock synchronization,
    it suffices to transform the estimators of $\alpha$, $\beta$, $\gamma$ and $\psi$ to 
    suitable estimators of $\rho$, $\freqd$ and $\slavephase$ through the comparison between 
    \eqref{eq:model} and \eqref{eq:stoch_model} (for details, see~\cite{AguilaPla2020a}).
    
    \subsection{Periodogram and correlation peaks (PCP), a fast and simple solution}
        
        Deliberately developed to be computationally cheap, PCP uses only very simple and efficient operations such as 
        discrete Fourier transforms (DFTs), sorting algorithms, and sample means. The estimator is divided in three 
        steps, and relies on the assumption that the sign of the amplitude $\psi$ is known. 
        First, one uses a periodogram of the $L$-$1$-times zero-padded centered data 
        to estimate the absolute value of the frequency parameter $\beta$, i.e., 
            $\hat{|\beta|} = \arg\max{k\in\mathcal{K}}{\left|\mathrm{DFT}_{NL}\left(\tilde{y}[n]\right)[k]\right|^2}/(NL)$
        where $\tilde{y}[n]$ is a length $NL$ signal such that
        \begin{IEEEeqnarray}{c} \label{eq:zero_padding}
            \tilde{y}[n] = \begin{cases}
                                y[n] - \frac{1}{N}\sum_{m=0}^{N-1} y[m] & \mbox{ if }n<N\mbox{, } \\
                                0 & \mbox{ if } N \leq n \leq NL-1\mbox{, } \nonumber
                           \end{cases}
        \end{IEEEeqnarray}
        and $\mathcal{K}=\lbrace 0, 1, \dots, \lfloor NL/2 \rfloor\rbrace$ is the set of indices representing the non-negative frequencies 
        in the DFT.  Note that, in this manner, $1/\hat{|\beta|}$ is a rough estimate of the period of the sawtooth signal.
        
        Second, one uses this unsigned frequency estimate to build two length $\lfloor 1/\hat{|\beta|}\rfloor$ signals
        $p_+[n]$ and $p_-[n]$ such that 
        $    p_{\pm}[n]  = \sign\left(\psi\right)\modu\left(\pm\hat{|\beta|}n \right)$
            for $0 \leq n < \lfloor1/\hat{|\beta|}\rfloor$.
        These two reference signals and the first estimated period of the data, i.e., the length $\lfloor 1/\hat{|\beta|} \rfloor$
        signal $\mathring{y}[n]$ such that $\mathring{y}[n] = y[n]$ for $0\leq n<1/\hat{|\beta|}$, are centered, max-normalized, and circularly
        correlated using length $\lfloor 1/\hat{|\beta|} \rfloor$ DFTs to estimate the sign of $\beta$ and the value of $\gamma$. In particular, if $\dot{y}[n]$, $\dot{p}_{+}[n]$,
        and $\dot{p}_-[n]$ are the centered and succesively max-normalized signals,
        one computes two numbers $l_+$ and $l_-$ as
            $l_\pm = \max{0 \leq n < 1/\hat{|\beta|}}{\mathrm{IDFT}\left[ 
             \mathrm{DFT}\left( \dot{p}_\pm[n]\right) \mathrm{DFT}\left(\dot{y}[n] \right)^* \right] }$, where $\cdot^*$ represents complex conjugation. Here, one also stores at which indices $n^{\mathrm{opt}}_\pm \in \listintzero{1/\hat{|\beta|}-1}$ the maxima $l_\pm$ are achieved. 
        Then, if $l_\pm>l_\mp$, one estimates $\hat{\beta}=\pm \hat{|\beta|}$ and $\hat{\gamma} = \modu\,(\hat{\beta}n^\mathrm{opt})$
        with $n^\mathrm{opt}=n^\mathrm{opt}_\pm$, and the amplitude of the signal is considered estimated as $\hat{\psi}_{\hat{\beta}}$
        through its relation with the frequency $\beta$.

        Third, one employs the closed-form solution for the minimum mean square error estimator for the offset parameter $\alpha$ assuming that $\hat{\beta}$,
        $\hat{\gamma}$ and $\hat{\psi}_{\hat{\beta}}$ are correct, i.e.,
        \begin{IEEEeqnarray}{c}\label{eq:concentrated_alpha}
            \hat{\alpha}_{\hat{\beta},\hat{\gamma}} = \sum_{n=0}^{N-1} y[n] 
                         - \sum_{m=0}^{N-1} \hat{\psi}_{\hat{\beta}} \modu\left[ \hat{\beta}m + \hat{\gamma}  \right]\mbox{.} 
        \end{IEEEeqnarray}
        Although this three-step estimator is heuristic, its computational 
        cost is very low, and it can be implemented in lightweight hardware. Furthermore, while some of its steps are rather 
        counter-intuitive, they show remarkable robustness. For example, using only the first estimated period of the data
        $\mathring{y}[n]$ to estimate the phase parameter $\gamma$ is clearly not an optimal strategy, but shows unparalleled robustness
        to errors in the estimation of the unsigned frequency parameter $|\beta|$, while steeply reducing the computational burden.
    
    \subsection{Local or global grid search (LGS or GGS), an exhaustive and costly solution} \label{sec:grid_search}
    
        \begin{figure}
        
            \input{figs/GGS_example_2}
            
            \vspace{-10pt}
            
            \caption{Example of the prediction mean squared error (PMSE, \eqref{eq:PMSE})
            of the model \eqref{eq:model} obtained in a global grid
            search procedure (as described in Section~\ref{sec:grid_search}) with a $10^3\times10^3$ grid with $\mathcal{B}=[0,10^{-2}]$ and 
            $\mathcal{G}=[0,1)$, when either $\beta$ or $\gamma$ are
            fixed to their approximated minimizing values $\beta^{\mathrm{opt}}$ and $\gamma^\mathrm{opt}$. 
            In this example, $N=500$, $\masterperiod = 10~\mathrm{ns}$, 
            $\samplingperiod = 100~\mu\mathrm{s}$, $\delta_0 = 5~\mu\mathrm{s}$, $\freqd = 73~\mathrm{Hz}$,
            $\slavephase = \pi~\mathrm{rad}$, and $\rho=2~\mathrm{m}$.
            For more details about the example and our implementation, as well as the image representation
            of the PMSE jointly over $\beta$ and $\gamma$, see this project's repository \cite{GITHUB}.
            \label{fig:GGS_example}}
        \end{figure}
        
        In contrast to PCP, the second technique we propose is computationally heavy. Nonetheless,
        our simulation study in Section~\ref{sec:numres} will suggest that it exhibits desirable statistical
        properties. In particular, we propose to minimize the prediction MSE (PMSE), i.e.,
        \begin{IEEEeqnarray}{c}\label{eq:PMSE}
            \!\!\!\!\!\!\underset{
                (\beta,\gamma)\in\mathcal{G}\times\mathcal{B}
            }{\operatorname{min}}\left\lbrace\!
            \sum_{n=0}^{N-1} \!\!\left( y[n] - \hat{\alpha}_{\beta,\gamma} -
            \hat{\psi}_{\beta} \modu\left[ \beta n + \gamma \right] \right)^2\!\right\rbrace\!\mbox{,}
        \end{IEEEeqnarray}
        In \eqref{eq:PMSE}, $\hat{\psi}_\beta$ is the implied estimator of $\psi$ for a given 
        $\beta$ we mentioned at the start of Section~\ref{sec:estimation}, and 
        $\hat{\alpha}_{\beta,\gamma}$ is the $\alpha$ that minimizes the cost function in \eqref{eq:PMSE},
        parametrized by $\beta$ and $\gamma$ and given \emph{mutatis mutandis} by the expression in 
        \eqref{eq:concentrated_alpha}.
        Regretfully, the solution to \eqref{eq:PMSE} has to be approximated, because the 
        PMSE over $\beta$ and $\gamma$ is neither convex nor unimodal, which implies that current iterative
        solvers are unable to find its global minimum efficiently. Example cuts of the profile of the PMSE over
        $\beta$ and $\gamma$ are reported in Fig.~\ref{fig:GGS_example}.
        We propose to approximately solve \eqref{eq:PMSE} by grid search, i.e., build a grid over some given
        ranges $\mathcal{G}\subset [0,1)$ for $\gamma$ and $\mathcal{B}\subset\left[-1/2,1/2\right)$ for 
        $\beta$ and pick the parameters $(\beta,\gamma)$ in the grid that yield the smallest value of
        the cost function. 
        We call this technique either global grid search (GGS) when $\mathcal{B}$ and $\mathcal{G}$ contemplate
        all possible values, and local grid search (LGS) when they are defined as small neighborhoods around the
        PCP estimates. The performance of these methods will critically depend on the number and location of the
        grid points, which are design parameters that set the compromise between accuracy and computational complexity.
        The simplest distribution of these grid points is uniformly accross $\mathcal{G}\times\mathcal{B}$, 
        with $N_\mathcal{G}$ possible values for $\gamma$ and $N_\mathcal{B}$ possible values for $\beta$.

    \section{Empirical results} \label{sec:numres}
        
        \begin{table}
            \caption{ Values for the parameters of PCP, LGS and GGS throughout 
                               the paper, unless otherwise stated. \label{tab:values_parameters_estimation} }
            \centering
            \begin{tabular}{c|c|c}
                    Parameter & Interpretation & Default value \\ \hline
                    $L$ & 
                    zero-padding factor & 
                    $5$ \\
                   $\mathcal{B}_{\mathrm{LGS}}$ & 
                    range for $\beta$ in LGS & 
                    $\hat{\beta}_\mathrm{PCP}+[-5,5]\cdot 10^{-4}$ \\
                    $\mathcal{G}_\mathrm{LGS}$ &
                    range for $\gamma$ in LGS & 
                    $\hat{\gamma}_\mathrm{PCP}+[-28,28]\cdot 10^{-3}$ \\
                    $(N_{\mathcal{B}},N_{\mathcal{G}})$ & 
                    gridpoints for LGS &
                    $(10^2, 10^3)$ \\
                    $\mathcal{B}_\mathrm{GGS}$ & 
                    range for $\beta$ in GGS & 
                    $[10^{-4},10^{-2}] $ \\
                    $\mathcal{G}_\mathrm{GGS}$ &
                    range for $\gamma$ in GGS &
                   $[0,1)$ \\
                    $(N_\mathcal{B},N_\mathcal{G})$ & 
                    gridpoints for GGS &
                    $(10^3, 10^3)$ \\
                    \hline
            \end{tabular}
        \end{table}

		\begin{figure}
			\centering        
			\input{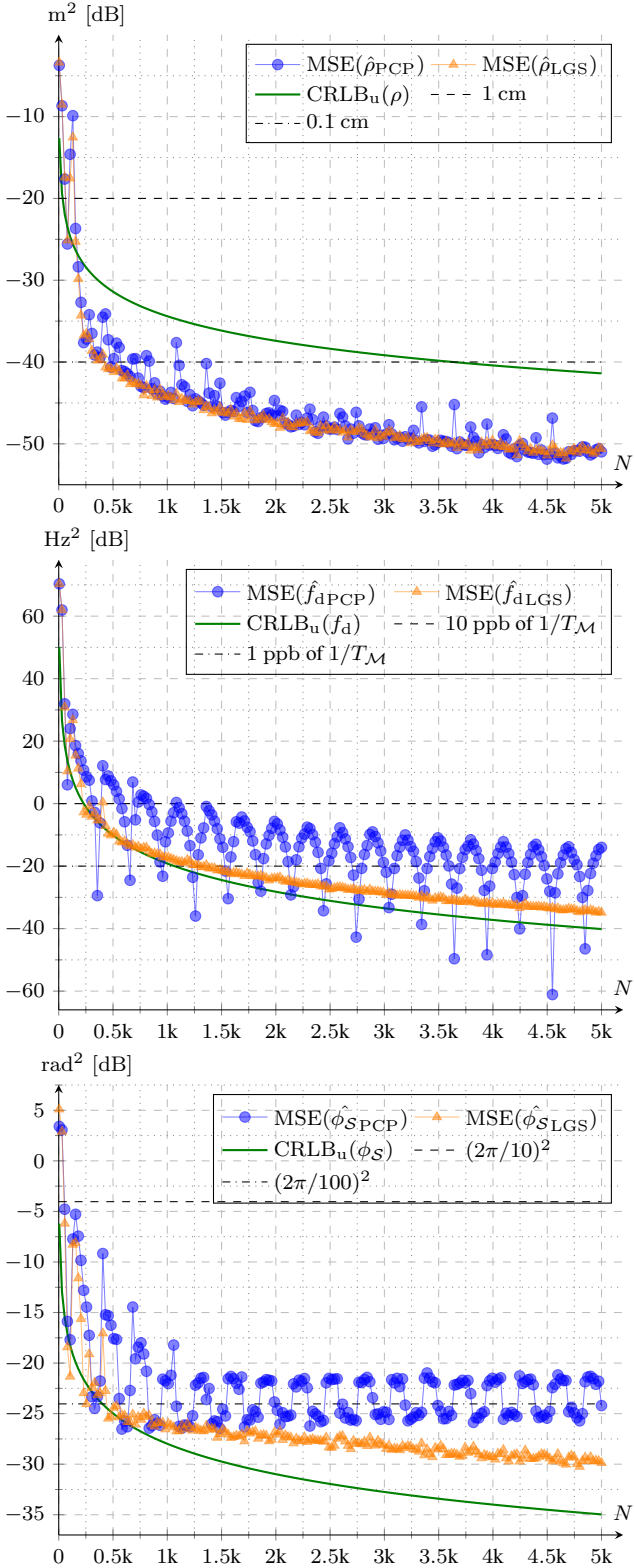}
			
			\vspace{-20pt}
			
			\caption{ Result of $300$ Monte Carlo repetitions for the physical parameters specified in Section~\ref{sec:numres}, evaluating the MSE in the estimation of
					$\rho$, $\freqd$ and $\slavephase$ by both PCP and LGS with respect to the 
					sample size. 
					For reference and comparison, we include the CRLBs for the unwrapped model
					derived in Section~\ref{sec:CRLBs}, and given by \eqref{eq:CRLB} and \eqref{eq:gradients}.
					\label{fig:vs_sample_size}}
		\end{figure}
	
		In Fig.~\ref{fig:vs_sample_size}, we illustrate the convergence of the $\mathrm{MSE}$ for the 
		PCP and LGS estimators proposed in Section~\ref{sec:estimation} with the sample size $N$ and 
		compare it with the CRLBs for the unwrapped model derived in \ref{sec:CRLBs}. The results we
		report were obtained from $300$ Monte Carlo repetitions for specific physical parameters, i.e., $\delay = 5~\mu\mathrm{s}$,
		$\masterperiod = 10~\mathrm{ns}$, 
		$\freqd=73~\mathrm{Hz}$, 
		$\samplingperiod=100~\mu\mathrm{s}$,
		$\rho=2~\mathrm{m}$, and
		$\slavephase=3\pi/4~\mathrm{rad}$. 
		Furthermore, the noise conditions were quite 
		benign 
		($\SNRin = 1/\sigma_v^2 = 40~\mathrm{dB}$
		and 
		$\SNRout = \Psi^2/\sigma_w^2 = 20~\mathrm{dB}$)
		and the algorithm's parameters were set as in Table~\ref{tab:values_parameters_estimation}.
		The results suggest that both estimators are consistent for these specific values of 
		the parameters, in the sense that their overall error tends to decrease with increasing sample 
		size, i.e., $\mathrm{MSE}\rightarrow 0$ with $N\rightarrow+\infty$. This is coherent with 
		the results we report in~\cite{AguilaPla2020a}, where we evaluate these algorithms
		with randomized physical parameters and under varying noise conditions.
		
		For PCP, the convergence of the MSE is clearly inefficient, and one observes it only by the decay of
		the envelope of the error. The regular bumps observed in the graphs of 
		$\mse{\hat{\freqd}_{\mathrm{PCP}}}$ and $\mse{\hat{\slavephase}_{\mathrm{PCP}}}$ are related to
		the resolution of the underlying periodogram estimate. On the one hand, if $\beta \approx k/NL$ for 
		some $k\in\listintzero{\lfloor NL/2\rfloor}$, $\beta$ will be included in the periodogram's
		grid and the PCP will be biased towards it and thus more likely to achieve very low MSE. On the other hand,
		if $\beta$ is between two such points, the PCP's bias will likely increase the MSE instead. 
		
		For LGS, the error seems to follow the decay of the CRLB of the unwrapped model in the estimation
		of $\rho$ and $\freqd$. However, the convergence of $\mse{\hat{\slavephase}_\mathrm{LGS}}$ is much
		slower than that predicted by the CRLB of the unwrapped model. This is to be expected, since the bounds
		in \eqref{eq:CRLB} do not take into account the non-linearity of the model, and therefore, the wrapping
		effect of the phase term. Although this non-linear behavior is what makes the joint estimation of 
		$\slavephase$ and $\rho$ possible, it also makes $\slavephase$ much harder to estimate than a simple offset.
		Furthermore, one must consider that the MSE in the estimation of $\slavephase$ only plays a role when one 
		aims to obtain time synchronization. If only phase synchronization is desired, however, consistence and 
		efficiency may be defined using more approriate evaluation metrics \cite[p.~84]{Mardia2009}.
		The evaluation with respect to these metrics is outside the scope of this paper. 
		
		For both PCP and LGS, the error in the estimation of the range $\rho$ is well below the CRLB, and 
		for $N\geq500$, it is mostly below $0.1~\mathrm{cm}$. Similarly, for PCP, $N\geq500$ leads to 
		average frequency estimation errors below $10~\mathrm{ppb}$ of $1/\masterperiod$ and average 
		phase estimation errors well below $2\pi/10$. For LGS, $N\geq500$ leads to average phase estimation
		errors below $2\pi/100$, and $N\geq1500$ to frequency estimation errors of less than $1~\mathrm{ppb}$
		of $1/\masterperiod$. 

		In conclusion, incorporating TDCs in wireless nodes to benefit from sawtooth modeling of 
		RTT measurements is a promising strategy to simultaneously achieve remarkable
		ranging and frequency synchronization accuracy (errors under $0.1~\mathrm{cm}$ and $1~\mathrm{ppb}$, respectively) and drastically decrease communication overhead. 
        On the other hand, absolute time synchronization seems to be less suited to the sawtooth 
        model, at least without more complex techniques (see the extended discussion we present
        in~\cite{AguilaPla2020a}).

    \section{Appendix: Fisher information matrix for the linear model with slope-dependent noise power} \label{sec:appFish}
          
  Consider the model for $\mathbf{Z}$ in \eqref{eq:vectorunwrappedmodel} and recall that 
  $\parsu=\trans{[\talpha,\tbeta]}$. \cite[ch.~3.9, p.~47]{Kay1993} provides the expression for the 
  Fisher information matrix of a generic Gaussian model in which $\mathbf{Z}\sim\normal{\mu_{\parsu}}{C_{\parsu}}$ as
  \begin{IEEEeqnarray*}{rl}
    \fish{}{\parsu} \,=
      \Bigg\lgroup &  \trans{\left[ \frac{\partial}{\partial \parsu_i}\mu_{\parsu}\right]} C_{\parsu}^{-1}  \left[ \frac{\partial}{\partial \parsu_j} \mu_{\parsu}\right] + \\
      & + \frac{1}{2}\tr\left[ C_{\parsu}^{-1} \frac{\partial C(\parsu)}{\partial \parsu_i} C_{\parsu}^{-1} \frac{\partial C(\parsu)}{\partial \parsu_j} \right] 
      \Bigg\rgroup_{i,j\in\lbrace1,2\rbrace}.
  \end{IEEEeqnarray*}
  For \eqref{eq:vectorunwrappedmodel}, $\partial \mu_{\parsu} / \partial \talpha = \ones{N}$, $\partial \mu_{\parsu} / \partial \tbeta = \n$, 
  $\partial C_{\parsu} / \partial \talpha = 0\, \Id{N}$ and 
  $\partial C_{\parsu} / \partial \tbeta = 2\sigma_2\left(\sigma_1 + \tbeta \sigma_2\right) \Id{N}$. Considering that 
  $\trans{\ones{N}} \ones{N}=N$, $\trans{\n} \ones{N} = N (N-1) / 2$ and $\trans{\n} \n = N(N-1)(2N-1)/6$, we obtain the Fisher information matrix 
  for \eqref{eq:vectorunwrappedmodel}, i.e., 
  \begin{IEEEeqnarray}{c}\label{eq:fish}
    \!\!\!\!\!\!\!\!\fish{}{\parsu} = \frac{N}{\sigma^2}\left(
      \begin{array}{cc}
      1 & \frac{N-1}{2} \\
      \frac{N-1}{2} & \frac{(N-1)(2N-1)}{6} + \frac{2\sigma_2^2\left(\sigma_1+\tbeta\sigma_2\right)^2}{\sigma^2}
      \end{array}\right).
  \end{IEEEeqnarray}
  Inverting  \eqref{eq:fish} leads to
  \begin{IEEEeqnarray}{c} \label{eq:invfish}
		\!\!\fish{-1}{\parsu} = \!
		\frac{\sigma^2 / N}{\frac{N+1}{12} + \frac{2 \sigma_2^2 \left( \sigma_1 + \tbeta \sigma_2 \right)^2}{\sigma^2 (N-1)}}
		\!\left(\!\!\! \begin{array}{cc}
		\frac{2N-1}{6} + \frac{2 \sigma_2^2 \left( \sigma_1 + \tbeta \sigma_2 \right)^2}{\sigma^2 (N-1)} \!\!& -\frac{1}{2} \\
		-\frac{1}{2} \!\! & \frac{1}{N-1}
		\end{array}\!\!
		\right)\!\! \mbox{ ,}\nonumber \\
	\end{IEEEeqnarray}
  which allows for the computation of the CRLBs for the estimation of $\talpha$ and $\tbeta$, and,
  through the relations \eqref{eq:relation_parameters_sync} and their
  gradients \eqref{eq:gradients}, the CRLBs for the estimation of $\freqd$, $\delaytrans$ when $\varphi_\slave$ is known, 
  and $\varphi_\slave$ when $\delaytrans$ is known. In terms of the rates of convergence for the variance of efficient estimators, 
  one can see that
  \begin{IEEEeqnarray}{rl}
		\fish{-1}{\parsu} & = \!\! \left( \begin{array}{cc}
								\!\!\frac{\sigma^2}{\frac{N(N+1)}{2(2N-1)} + \mathcal{O}(N^{-1})} + \frac{2 \sigma_2^2 (\sigma_1 + \tilde{\beta} \sigma_2)^2}{\frac{N(N^2-1)}{12} + \mathcal{O}(1) } \!\!\!\!&
								- \frac{\sigma^2}{\frac{N (N+1)}{6} + \mathcal{O}(1)} \\
								\!\!- \frac{\sigma^2}{\frac{N (N+1)}{6} + \mathcal{O}(1)} \!\!\!\!&
								\frac{\sigma^2}{\frac{N(N^2-1)}{12} + \mathcal{O}(N)}
		           \end{array} \right)\!\!, \nonumber
  \end{IEEEeqnarray}
  i.e., the efficient estimators of the offset $\tilde{\alpha}$ and the slope $\tilde{\beta}$ still have the same rates of convergence as in a standard linear model,
  with additions of only non-dominating terms.

\bibliographystyle{IEEEbib}
\bibliography{clock-sync}

\end{document}